\documentclass[conference]{IEEEtran}
\pdfoutput=1
\IEEEoverridecommandlockouts

\usepackage{acronym}
\usepackage{amsmath,amssymb}
\usepackage[english]{babel}
\usepackage{bbm}
\usepackage{bm}
\usepackage{caption}
\usepackage{cite}
\usepackage{enumerate}
\usepackage[T1]{fontenc}
\usepackage{graphicx}
\usepackage{hyperref}
\usepackage{cleveref}
\usepackage[utf8]{inputenc}
\usepackage{mathtools}
\usepackage{siunitx}
\usepackage{subcaption}
\usepackage{xcolor}
\usepackage{algorithm}
\usepackage{algorithmic}

\definecolor{myblue}{RGB}{31,119,180}
\definecolor{myorange}{RGB}{255,127,14}
\definecolor{mygreen}{RGB}{44,160,44}

\acrodef{CME}{chemical master equation}
\acrodef{PBS}{particle-based computer simulation}
\acrodef{NT}{neurotransmitter}
\acrodef{LNA}{linear noise approximation}
\acrodef{MC}{molecular communication}
\acrodef{DMC}{diffusive molecular communication}

\newcommand{\kd}{\ensuremath{\kappa_\mathrm{d}}}
\newcommand{\ke}{\ensuremath{\kappa_\mathrm{e}}}
\newcommand{\ka}{\ensuremath{\kappa_\mathrm{a}}}
\renewcommand{\P}{\ensuremath{\textrm{P}}}
\newcommand{\dt}{\ensuremath{\mathrm{d}t}}
\newcommand{\dtau}{\ensuremath{\mathrm{d}\tau}}
\newcommand{\kadt}{\ensuremath{\tilde{\kappa}_{a_0}}}
\newcommand{\pibd}{\ensuremath{\boldsymbol{\pi}}}
\newcommand{\Abd}{\ensuremath{\boldsymbol{A}}}
\newcommand{\Qbd}{\ensuremath{\boldsymbol{Q}}}
\newcommand{\Dbd}{\ensuremath{\boldsymbol{D}}}
\newcommand{\Nmax}{\ensuremath{N_\textrm{max}}}
\newcommand{\Nmin}{\ensuremath{N_\textrm{min}}}
\newcommand{\Omax}{\ensuremath{O_\textrm{max}}}
\newcommand{\Omin}{\ensuremath{O_\textrm{min}}}

\newtheorem{theorem}{Theorem}

\allowdisplaybreaks

\renewcommand{\baselinestretch}{0.93}

\makeatletter
\long\def\@makecaption#1#2{\ifx\@captype\@IEEEtablestring%
    \footnotesize\begin{center}{\normalfont\footnotesize #1}\\
        {\normalfont\footnotesize\scshape #2}\end{center}%
    \@IEEEtablecaptionsepspace
    \else
    \@IEEEfigurecaptionsepspace
    \setbox\@tempboxa\hbox{\normalfont\footnotesize {#1.}~~ #2}%
    \ifdim \wd\@tempboxa >\hsize%
    \setbox\@tempboxa\hbox{\normalfont\footnotesize {#1.}~~ }%
    \parbox[t]{\hsize}{\normalfont\footnotesize \noindent\unhbox\@tempboxa#2}%
    \else
    \hbox to\hsize{\normalfont\footnotesize\hfil\box\@tempboxa\hfil}\fi\fi}
\makeatother

\begin{document}
    
\setlength{\abovedisplayskip}{3pt}
\setlength{\belowdisplayskip}{3pt}
\setlength{\textfloatsep}{2pt}
    
\title{A Chemical Master Equation Model for Synaptic Molecular Communication
    \thanks{This work was supported in part by the German Research Foundation (DFG) under grant SCHO 831/9-1.}
    \vspace{-2mm}
}

\author{\IEEEauthorblockN{Sebastian Lotter, Maximilian Sch\"afer, and Robert Schober\\} 
    \IEEEauthorblockA{Friedrich-Alexander University Erlangen-N\"urnberg, Germany}
    \vspace*{-5mm}
}
\maketitle

\begin{abstract}
    In synaptic molecular communication, the activation of postsynaptic receptors by \acp{NT} is governed by a stochastic reaction-diffusion process and, hence, inherently random.
    It is currently not fully understood how this randomness impacts downstream signaling in the target cell and, ultimately, neural computation and learning.
    The statistical characterization of the reaction-diffusion process is difficult because the reversible bi-molecular reaction of \acp{NT} and receptors renders the system nonlinear.
    Consequently, existing models for the receptor occupancy in the synaptic cleft rely on simplifying assumptions and approximations which limit their practical applicability.
    In this work, we propose a novel statistical model for the reaction-diffusion process governing synaptic signal transmission in terms of the \ac{CME}.
    We show how to compute the \ac{CME} efficiently and verify the accuracy of the obtained results with stochastic \acp{PBS}.
    Furthermore, we compare the proposed model to two benchmark models proposed in the literature and show that it provides more accurate results when compared to \acp{PBS}.
    Finally, the proposed model is used to study the impact of the system parameters on the statistical dependence between binding events of \acp{NT} and receptors.
    In summary, the proposed model provides a step forward towards a complete statistical characterization of synaptic signal transmission.
\end{abstract}
\acresetall

\section{Introduction}
\label{sec:introduction}
\Ac{DMC} is a novel communication paradigm inspired by the exchange of information between biological entities by means of diffusing molecules\cite{nakano13}.
It is envisioned that \ac{DMC} will enable revolutionary applications in the field of intra-body nano-scale communications based on and interfacing with natural \ac{MC} systems, such as the synaptic \ac{DMC} system\cite{veletic2019}.
Since the synaptic \ac{DMC} system enables complex processes such as learning and memory, understanding its design is key to the development of synthetic neural applications such as neural prostheses and brain-machine interfaces\cite{veletic2019}.
However, despite considerable research efforts, our picture of synaptic communication is not complete to date.

In synaptic \ac{DMC}, information is conveyed from the {\em presynaptic} cell to the {\em postsynaptic} cell by means of diffusing molecules called \acp{NT}.
\acp{NT} are sensed by the postsynaptic cell using membrane-bound receptors and may be degraded by enzymes while diffusing in the extracellular medium\cite{zucker14}.
One central open question regarding synaptic \ac{DMC} concerns the role of different kinds of signaling noise for the signal transmission\cite{veletic2019}.
In particular, it is not fully understood, whether the randomness in synaptic communication due to the random propagation and reaction of \acp{NT} with enzymes and postsynaptic receptors makes synaptic communication more or less reliable\cite{rusakov20}.

Synaptic \ac{DMC} has been studied in the \ac{MC} community with emphasis on different aspects, such as information theoretic limits \cite{veletic20}, the design of artificial synapses \cite{bilgin17}, and the long-term average signal decay \cite{oncu21}, see also literature overviews in \cite{veletic2019,lotter20}.
Mean-field models, i.e., deterministic models for the average receptor activation valid in the large system limit, have been developed for synapses employing enzymatic degradation \cite{lotter21,oncu21} and other channel clearance mechanisms \cite{khan2017,lotter20,bilgin17}.
However, stochastic fluctuations in the postsynaptic receptor activation have been considered only recently \cite{lotter21}.
Yet, the statistical model proposed in \cite{lotter21} does not account for the randomness of the enzymatic degradation of \acp{NT} and relies on the simplifying assumption that either \acp{NT} compete for postsynaptic receptors or postsynaptic receptors compete for \acp{NT}.
Hence, the scope and applicability of the model in \cite{lotter21} is limited to a specific range of parameter values.
Other statistical models for ligand receptors in the \ac{MC} literature assume statistical independence of the receptors \cite{kuscu18}.
As already shown in \cite{lotter21}, this assumption is in general not justified.

In this paper, we propose a novel statistical signal model for synaptic \ac{DMC} in terms of the \ac{CME}.
The proposed model characterizes the joint statistics of the postsynaptic receptor occupancy and the enzymatic degradation process for the first time in the \ac{MC} literature.
Furthermore, in contrast to existing models, it does not rely on simplifying assumptions with respect to the statistical (in)dependence between receptors and/or \acp{NT}.
Since the \ac{CME} model in its original form is computationally intractable, a novel adaptive state reduction scheme is proposed which allows the efficient computation of the proposed model.
The proposed state reduction scheme exploits knowledge of the first-order statistics of the considered process and, in contrast to common approximation methods for the \ac{CME} found in the literature\cite{munsky18}, the approximation error is explicitly characterized and can, hence, be controlled.
Finally, the results of the proposed model are compared to stochastic \acp{PBS} to verify both the assumptions made to arrive at the proposed \ac{CME} model and the accuracy of the proposed state reduction scheme.
In summary, the proposed model allows for an accurate statistical characterization of the synaptic noise due to \ac{NT} binding and degradation outperforming existing models.
It hence provides a step forward towards a comprehensive statistical analysis of the noise in synaptic \ac{DMC}.

The remainder of this paper is organized as follows.
The system model in terms of the \ac{CME} is introduced in Section~\ref{sec:system_model}.
In Section~\ref{sec:cme}, a state reduction scheme for the computation of the \ac{CME} is provided.
In Section~\ref{sec:numerical_results}, the proposed model is used to study the statistics of synaptic communication for selected, biologically relevant parameter regimes and numerical results from \ac{PBS} are presented which validate the model.
Section~\ref{sec:conclusion} concludes the paper with a brief summary of the main findings and an outlook towards future research directions.

\section{System Model}
\label{sec:system_model}
\begin{figure}[!tbp]
    \centering
    \includegraphics[width=.4\textwidth]{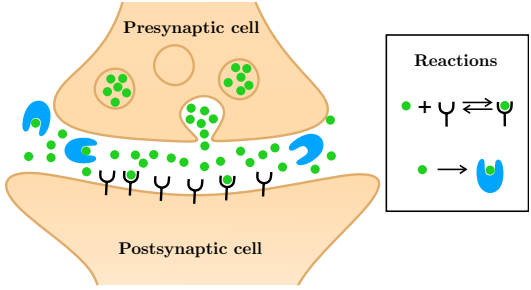}
    \caption{Chemical synapse. \acp{NT} (green) are released by exocytosis at the presynaptic cell and diffuse in the synaptic cleft. Solute \acp{NT} can bind reversibly to receptors (black) at the postsynaptic membrane and may be degraded by enzymes (blue). The two reactions considered for the statistical analysis in this paper are the reversible bimolecular reaction of \acp{NT} with postsynaptic receptors and the unimolecular degradation process modeling the degradation of \acp{NT} by enzymes.}
    \label{fig:synapse}
\end{figure}

In synaptic communication, \acp{NT} are released into the synaptic cleft by exocytosis of presynaptic vesicles.
After release, \acp{NT} propagate by Brownian motion and react with postsynaptic transmembrane receptors and degradative enzymes, cf.~Fig.~\ref{fig:synapse}.
Since both reactions and the diffusion of \acp{NT} are random, the number of activated postsynaptic receptors at time $t$, $O(t)$, and the total number of (non-degraded) \acp{NT} at time $t$, $N(t)$, are random processes.

\subsection{Mean-Field Model}

For synaptic \ac{DMC} systems satisfying the assumptions discussed in \cite[Sec.~II-A]{lotter21}, i.e., synapses that employ enzymatic degradation as channel clearance mechanism and are either of large extent or confined by surrounding cells, the {\em expected} concentration of \acp{NT} after a single release into the synaptic cleft is described by the following partial differential equation \cite{lotter21}
\begin{align}
    \partial_t c(x,t) = D\partial_{xx} c(x,t) - \ke c(x,t), \quad 0 < x < a,\label{eq:mean:rd}
\end{align}
where the synaptic cleft is represented by the one-dimensional spatial domain $[0,a]$, $c(x,t)$ denotes the expected concentration of solute \acp{NT} at time $t$ and spatial coordinate $x$ in $\si{\per\micro\meter}$, $D$ and $\ke$ denote the diffusion coefficient for \acp{NT} in $\si{\micro\meter\squared\per\micro\second}$ and the enzymatic degradation rate in $\si{\per\micro\second}$, respectively, and $\partial_t$ and $\partial_{xx}$ denote the first partial derivative with respect to time $t$ and the second partial derivative with respect to space $x$, respectively.
The reversible binding of \acp{NT} to postsynaptic receptors is modeled as a boundary condition at $x=a$ \cite{lotter21}
\begin{equation}
    - D\,\partial_x c(x,t)\big\vert_{x=a} = \ka \left( 1 - \frac{o(t)}{C} \right) c(a,t) - \kd o(t),\label{eq:mean:sat_bdr}
\end{equation}
where $C$ and $o(t)$ denote the total number of postsynaptic receptors and the {\em expected} number of postsynaptic receptors occupied at time $t$, respectively, $\ka$ and $\kd$ denote the microscopic binding rate of \acp{NT} to postsynaptic receptors in $\si{\micro\meter\per\micro\second}$ and the unbinding rate of \acp{NT} from postsynaptic receptors in $\si{\per\micro\second}$, respectively, and $\partial_x$ denotes the first partial derivative with respect to $x$.
The model is completed by the initial and boundary conditions \cite{lotter21}
\begin{align}
    c(x,0) = N_0\delta(x) \quad\textrm{and}\quad \partial_x c(x,t)\big\vert_{x=0} = 0,\label{eq:mean:init_and_no_flux_bdr}
\end{align}
respectively, where $N_0$ and $\delta(\cdot)$ denote the number of released \acp{NT} and the Dirac delta distribution, respectively.
Furthermore, $o(t)$ is related to $c(x,t)$ by the equation
\begin{equation}
    o(t) = \int_{0}^{t} - D\,\partial_x c(x,\tau)\big\vert_{x=a} \dtau.\label{eq:mean:def_i}
\end{equation}

Since boundary condition \eqref{eq:mean:sat_bdr} is nonlinear, a closed-form solution to the boundary value problem \labelcref{eq:mean:rd,eq:mean:sat_bdr,eq:mean:init_and_no_flux_bdr} cannot be obtained.
Instead, a state space model is used in \cite{lotter21} to compute $o(t)$ iteratively in the transfer function domain.
We call this model $\mathcal{S}$ and it is defined by a {\em state equation} \cite[Eq.~(42)]{lotter21} and an {\em output equation} \cite[Eq.~(31)]{lotter21}.

Now, let $n(t)$ denote the expected total number of \acp{NT}, i.e., the expected number of solute \acp{NT} and bound \acp{NT}, at time $t$.
When $\mathcal{S}$ is computed, we obtain not only $o(t)$, but also $c(x,t)$ and $n(t)$ \cite[Sec.~III-C-3]{lotter21}.

In the following, we will use these quantities to compute the macroscopic absorption rate for \acp{NT} to postsynaptic receptors.

\subsection{Macroscopic Binding Rate}

In the mean-field model \labelcref{eq:mean:rd,eq:mean:sat_bdr,eq:mean:init_and_no_flux_bdr,eq:mean:def_i}, the binding rate of \acp{NT} to postsynaptic receptors is given by constant $\ka$.
In fact, $\ka$ results from mapping the actual three-dimensional reaction-diffusion process to the one-dimensional process in \labelcref{eq:mean:rd,eq:mean:sat_bdr,eq:mean:init_and_no_flux_bdr}\footnote{By ``one-dimensional'', we refer to spatial dimensions, excluding the temporal dimension.}.
According to \cite[Sec.~V-A]{lotter21}, $\ka$ can be written as
\begin{equation}
    \ka = \kadt C,
\end{equation}
where $\kadt$ is a constant depending on the intrinsic binding rate of one \ac{NT} to one receptor and the ratio of the receptor area to the postsynaptic membrane surface area.
Hence, the activation of postsynaptic receptors can be written as the following reversible bi-molecular reaction \eqref{eq:mean:sat_bdr}
\begin{equation}
    S_a + R \xrightleftharpoons[\kd]{\kadt} O,\label{eq:reac:intr_bd_rt}
\end{equation}
where $R$ denotes unoccupied postsynaptic receptors and $S_a$ denotes the solute \acp{NT} located in an (infinitesimally) small volume close to the postsynaptic membrane.

Now, denoting by $S(t)$ the total number of solute molecules at time $t$ and assuming that the ratio $S_a(t)/S(t)$ is well-approximated by the ratio of the corresponding mean values obtained from \labelcref{eq:mean:rd,eq:mean:sat_bdr,eq:mean:init_and_no_flux_bdr,eq:mean:def_i}, i.e.,
\begin{align}
    \frac{S_a(t)}{S(t)} \approx \frac{\mathbb{E}[S_a(t)]}{\mathbb{E}[S(t)]} = \frac{c(a,t)}{\int_{0}^{a}c(x,t)\mathrm{d}x},\label{eq:diff_approx}
\end{align}
where $\mathbb{E}[\cdot]$ denotes the expectation operator, we can write the change in $O$ due to the binding and unbinding of \acp{NT} in the large system limit as
\begin{equation}
\frac{\mathrm{d}O(t)}{\mathrm{d}t} = \kadt S(t) \frac{c(a,t)}{\int_{0}^{a}c(x,t)\mathrm{d}x} R(t) - \kd O(t).
\end{equation}
Hence, defining the time-dependent macroscopic binding rate $\ka(t)$ as $\ka(t) = \kadt c(a,t)/\int_{0}^{a}c(x,t)\mathrm{d}x$, we obtain the reaction
\begin{equation}
    S + R \xrightleftharpoons[\kd]{\ka(t)} O.\label{eq:reac:macr_bd_rt}
\end{equation}

Eq.~\eqref{eq:reac:macr_bd_rt} provides a space-independent description of the reaction of \acp{NT} with postsynaptic receptors.
However, in contrast to space-independent models with constant reaction rates \cite{munsky18}, we do {\em not} assume that the reaction volume is well-mixed.
Instead, the spatially heterogeneous and time-dependent distribution of solute \acp{NT} is represented by $\ka(t)$.
The accuracy of this model as compared to the actual reaction-diffusion process depends on the validity of \eqref{eq:diff_approx}.
Eq.~\eqref{eq:diff_approx} in turn is justified if the number of solute \acp{NT} is not too small and diffusion is relatively fast as compared to the chemical reactions.
We will see in Section~\ref{sec:numerical_results} that \eqref{eq:reac:macr_bd_rt} provides a very accurate model for different, biologically relevant ranges of parameter values.

\subsection{The Chemical Master Equation}

In this section, we formulate a statistical model for the random processes $O(t)$ and $N(t)$ in terms of the \ac{CME}.

First, besides the reaction of \acp{NT} with postsynaptic receptors defined in \eqref{eq:reac:macr_bd_rt}, solute \acp{NT} are exposed to enzymatic degradation which is modeled as a uni-molecular reaction in \eqref{eq:mean:rd}.
This degradation reaction is described as follows
\begin{align}
    S \xrightharpoonup{\ke} \varnothing,\label{eq:reac:enz_deg}
\end{align}
where $\varnothing$ denotes any species that does not react with \acp{NT} and postsynaptic receptors.

Next, we note that the state of the system described by \labelcref{eq:reac:macr_bd_rt,eq:reac:enz_deg} at time $t$ is fully determined by the random variables $N(t)$ and $O(t)$, since $S(t) = N(t) - O(t)$ and $R(t) = C - O(t)$.
Further, denoting the time-dependent joint probability mass function of $N(t)$ and $O(t)$ as $\P(n,o,t)$, i.e., $\P(n,o,t) = \Pr\lbrace (N(t),O(t))=(n,o)\rbrace$, the time-evolution of $\P(n,o,t)$ is governed by the \ac{CME} \eqref{eq:cme}, shown at the top of the next page,
\begin{figure*}
    \begin{align}
    \frac{\partial \P(n, o, t)}{\partial t} = &-\left[\kd o + \ke (n-o) + \ka(t)(n-o)(C-o)\right]\P(n,o,t) + \kd(o+1)\P(n,o+1,t) \nonumber\\
    &{} + \ke (n + 1 - o) \P(n+1,o,t) + \ka(t) (n-o+1)(C-o+1)\P(n,o-1,t)\label{eq:cme}
    \end{align}
    \vspace*{-9mm}
\end{figure*}
where $n \in \left\lbrace 0,\ldots,N_0 \right\rbrace$, $o \in \left\lbrace 0,\ldots,C \right\rbrace$.
To define \eqref{eq:cme} on the boundary of the state space, we set $\P(-1, \cdot, \cdot) \equiv 0$, $\P(N_0+1, \cdot, \cdot) \equiv 0$, $\P(\cdot, -1, \cdot) \equiv 0$, and $\P(\cdot, C+1, \cdot) \equiv 0$.

The \ac{CME} \eqref{eq:cme} defines a discrete-state random process whose state transition probabilities are determined by \eqref{eq:cme}.
This process is not Markovian, because the state-transition probabilities are time-dependent.

\section{Solving the Chemical Master Equation}    
\label{sec:cme}
A closed-form solution of the system of equations \eqref{eq:cme} is in general not possible \cite{schnoerr17}.
Hence, in this section, we aim at computing $P(n,o,t)$ as defined in \eqref{eq:cme} numerically.
As we will see, even the numerical evaluation of \eqref{eq:cme} poses a severe challenge.

\subsection{Formal Solution}

According to \eqref{eq:cme}, there exist $M=(N_0+1)\times(C+1)$ different system states.
We organize these states in a level-dependent manner where the total number of \acp{NT}, $n$, determines the level.
This means, we define the probability vector $\pibd(t) \in [0,1]^{M}$ as
\begin{equation}
    \pibd(t) = [\pibd_{N_0}(t), \pibd_{N_0-1}(t), \ldots, \pibd_{0}(t)]^{\mathrm{T}},\label{eq:def:pibd}
\end{equation}
where $[\cdot]^{\mathrm{T}}$ denotes transposition and the $N_0+1$ vectors $\pibd_{n}(t) \in [0,1]^{(C+1)}$ are defined as\footnote{Note that we allow for infeasible states in this definition, since $P(n,o,\cdot) \equiv 0$ for $n < o$. This is done only for notational simplicity, infeasible states are omitted in all practical computations.} $\pibd_{n}(t) = [P(n,0,t),P(n,1,t),\ldots,P(n,C,t)]$.

In a similar fashion, we collect all transition probabilities from \eqref{eq:cme} in the time-dependent transition matrix $\Abd(t) \in \mathbb{R}^{M \times M}$.
$\Abd(t)$ is a block-bidiagonal matrix consisting of $(N_0+1)^2$ $(C+1) \times (C+1)$ matrices with all sub-matrices equal to the quadratic all-zero matrix $\boldsymbol{0}_{(C+1)}$, except for the matrices on the main diagonal and the lower diagonal which we denote as
\begin{equation}
    \Abd_{i,i} = \Qbd_{N_0-i+1},\quad 1 \leq i \leq N_0+1,\label{eq:A:Q}
\end{equation}
and
\begin{equation}
    \Abd_{i,i-1} = \Dbd_{N_0-i+1},\quad 2 \leq i \leq N_0+1,\label{eq:A:D}
\end{equation}
respectively.
Matrices $\Qbd_{n}$ and $\Dbd_{n}$ collect the level-dependent transition rates for the binding and the degradation reactions.
The $\Qbd_{n}$ are tridiagonal matrices with the diagonals defined as follows
\begin{align}
    &\left(\Qbd_{n}\right)_{i+1,i+1}&=\,& -\left[\kd i + \ke (n-i) \right.\nonumber\\
    &&&\left.+ \ka(t)(n-i)(C-i)\right],\qquad\, 0 \leq i \leq C,\\
    &\left(\Qbd_{n}\right)_{i+1,i}&=\,& \ka(t) (n-i+1)(C-i+1),\,\, 1 \leq i \leq C,\\
    &\left(\Qbd_{n}\right)_{i+1,i+2}&=\,& \kd (i+1),\qquad\,\, 0 \leq i \leq C-1.
\end{align}
The $\Dbd_{n}$ are diagonal matrices with diagonals defined as follows
\begin{equation}
    \left(\Dbd_{n}\right)_{i+1,i+1} = \ke (n + 1 - i),\qquad\qquad 0 \leq i \leq C.
\end{equation}
With these definitions, we can write \eqref{eq:cme} in vector form as the following system of differential equations
\begin{equation}
    \frac{\mathrm{d}\pibd}{\dt} = \Abd(t)\pibd(t),\label{eq:cme:vec}
\end{equation}
the formal solution of which is given as
\begin{equation}
    \pibd(t) = \exp\left(\int_0^t \Abd(\tau) \dtau\right)\pibd_0,\label{eq:cme:vec:sol}
\end{equation}
where $\exp(\boldsymbol{M})$ denotes the matrix exponential of square matrix $\boldsymbol{M}$ and, according to \eqref{eq:mean:init_and_no_flux_bdr}, $\pibd_0 = \pibd(0)$ is given by the $M$-dimensional vector $[1,0,\ldots,0]^{\mathrm{T}}$.

\subsection{Computational Issues and Approximation Schemes}

Since the dimension of $\Abd(t)$ grows quadratically with both the number of released \acp{NT} and the number of receptors, computing the matrix exponential in \eqref{eq:cme:vec:sol} is intractable \cite{moler03}.
Indeed, even for the moderate number of $500$ released \acp{NT} and $200$ receptors, the number of elements of $\Abd(t)$ is of order $\sim 10^{10}$.

This problem is common to many applications using the \ac{CME} as modeling tool and, consequently, several methods have been proposed to approximate the solution of the \ac{CME} \cite{schnoerr17}.
Two of the most frequently used approximation schemes are {\em moment closure} schemes and schemes exploiting some kind of {\em system size expansion}, the most popular among the latter being the {\em \ac{LNA}} \cite{schnoerr17}.
Both of these approaches have their strengths and limitations, the discussion of which would go far beyond the scope of this paper.
Here, it suffices to say that due to the bimolecular reaction \eqref{eq:reac:macr_bd_rt} both methods cannot be used without further simplifications or approximations to obtain the statistics of $N(t)$ and $O(t)$.

Another commonly used method for computing high-dimensional \acp{CME} is to approximate the \ac{CME} on a lower-dimensional subspace of its state space\footnote{Such state reduction schemes are also referred to as {\em state lumping schemes} \cite{munsky18}.}.
Classical state reduction schemes for the \ac{CME} operate on a reduced but static state space, meaning the state space does not change over time \cite{munsky18}.
In the following section, we show how to exploit our knowledge of the first-order statistics of $N(t)$ and $O(t)$ given by $n(t)$ and $o(t)$, respectively, to adapt the state space iteratively while computing the \ac{CME}.
We show that this adaptive scheme allows to compute \eqref{eq:cme:vec:sol} efficiently and, at the same time, control the approximation error.

\vspace*{-1mm}
\subsection{Adaptive State Reduction}\label{sec:cme:state_reduction}

To introduce the proposed adaptive state reduction scheme, we first discretize time into subsequent intervals of length $\Delta t$, such that the $k$th interval is $I_k = [t_k,t_{k+1}]$, where $t_k = (k-1) \Delta t$ and $k$ is from the set of positive integers $\mathbb{N}$.
The idea is to compute $\pibd(t)$ iteratively for each interval $k$ while discarding the states $(n,o)$ which do not contribute significant probability mass in interval $k$.

To this end, we first define the respective marginal distributions of $N(t)$ and $O(t)$ at time $t$ as
\begin{equation}
    P_N(n,t) = \sum_{o=0}^{C} P(n,o,t) \,\,\textrm{and}\,\, P_O(o,t) = \sum_{n=0}^{N_0} P(n,o,t),
\end{equation}
and the {\em full state space} of \eqref{eq:cme} as
\begin{equation}
    S_0 = \lbrace (n,o)| 0 \leq n \leq N_0, 0 \leq o \leq C \rbrace.\label{eq:def:S_0}
\end{equation}
Furthermore, let $P_B(\cdot;n,p)$ denote the probability mass function of a binomial random variable with parameters $n$ and $p$, $\epsilon > 0$, and define
\begin{align}
    \Nmin^{(k)} &= \max \left\lbrace n \left| \sum_{n'=0}^{n} P_B\left(n';N_0,\frac{n(t_{k+1})}{N_0}\right)\right. < \epsilon \right\rbrace,\label{eq:def:Nmin}\\
    \Nmax^{(k)} &= \min \left\lbrace n\, \left| \sum_{n'=n}^{N} P_N\left(n',t_k\right) < \epsilon \right\rbrace\right.,\label{eq:def:Nmax}\\
    \Omin^{(k)} &= \max \left\lbrace o\, \left| \max_{t \in I_k} \sum_{o'=0}^{o} P_B\left(o';C,\frac{i(t)}{C}\right) < \epsilon \right\rbrace\right.,\label{eq:def:Omin}\\
    \Omax^{(k)} &= \min \left\lbrace o\, \left| \max_{t \in I_k} \sum_{o'=o}^{C} P_B\left(o';C,\frac{i(t)}{C}\right) < \epsilon \right\rbrace\right..\label{eq:def:Omax}
\end{align}
We define the {\em reduced state space} in interval $k$ as follows
\begin{equation}
    S_k = \lbrace (n,o)| \Nmin^{(k)} \leq n \leq \Nmax^{(k)}, \Omin^{(k)} \leq o \leq \Omax^{(k)} \rbrace,\label{eq:def:S_k}
\end{equation}
and the restriction of $\pibd(t)$ to $S_k$ as
\begin{equation}
    \pibd(t)|_{S_k} = [\pibd_{(\Nmax^{(k)})}(t)|_{S_k}, \ldots, \pibd_{(\Nmin^{(k)})}(t)|_{S_k}]^{\mathrm{T}},\label{eq:def:pibd_S_k}
\end{equation}
where
\begin{equation}
    \pibd_{n}(t)|_{S_k} = [P(n,\Omin^{(k)},t),\ldots,P(n,\Omax^{(k)},t)].
\end{equation}
The restriction of $\Abd(t)$ to $S_k$, $\Abd(t)|_{S_k}$, is obtained by discarding the rows and the columns of $\Abd(t)$ corresponding to the indices of the elements of $\pibd(t)$ discarded in $\pibd(t)|_{S_k}$.
Finally, we define the approximate solution of \eqref{eq:cme:vec} in interval $I_k$, $\hat{\pibd}^{(k)}(t)$, as the solution of the following system of equations
\vspace*{-1mm}
\begin{equation}
    \frac{\mathrm{d} \hat{\pibd}^{(k)}(t)}{\mathrm{d} t} = \Abd(t)|_{S_k} \hat{\pibd}^{(k)}(t),\label{eq:def:pibd_hat}
\end{equation} 
where $t \in I_k$ and $\hat{\pibd}^{(k)}(t_k) = \pibd(t_k)|_{S_k}$.

The following theorem justifies these definitions.
\begin{theorem}\label{thm:state_reduction}
    Let $k \in \mathbb{N}$ and assume $\pibd(t_k)$ is known. Then, for any $\epsilon > 0$,
    \begin{equation}
        ||\pibd(t)|_{S_k} - \hat{\pibd}^{(k)}(t)||_1 < 4 \epsilon, \quad\forall\, t \in I_k,
    \end{equation}
    where $\pibd(t)|_{S_k}$, $S_k$, and $\hat{\pibd}^{(k)}(t)$ are defined in \eqref{eq:def:pibd_S_k}, \eqref{eq:def:S_k}, and \eqref{eq:def:pibd_hat}, respectively, and $||\boldsymbol{v}||_1$ denotes the $l_1$ norm of vector $\boldsymbol{v}$\footnote{State space $S_k$ depends on the choice of $\epsilon$ by definitions \labelcref{eq:def:Nmin,eq:def:Nmax,eq:def:Omin,eq:def:Omax,eq:def:S_k}. This dependence remains implicit for notational simplicity.}.
\end{theorem}
\begin{IEEEproof}
    See Appendix~\ref{sec:app:prf_thm_1}.
\end{IEEEproof}

Thm.~\ref{thm:state_reduction} allows us to approximate the solution to the \ac{CME} \eqref{eq:cme} by iteratively solving the lower-dimensional problem \eqref{eq:def:pibd_hat} for each interval $k$.
To state the iterative algorithm, we need yet to define how to map $\hat{\pibd}^{(k)}(t)$ to $\hat{\pibd}^{(l)}(t)$ for any $k,l \in \mathbb{N} \bigcup \lbrace 0 \rbrace$.
To this end, let us denote the elements of $\hat{\pibd}^{(k)}(t)$ by
\begin{multline}
    \hat{\pibd}^{(k)}(t) =\\
    [\hat{P}^{(k)}(n_{i_1},o_{i_1},t),\ldots,\hat{P}^{(k)}(n_{i_{|S_{k}|}},o_{i_{|S_{k}|}},t)]^{\mathrm{T}},
\end{multline}
where the indices $i_1,\ldots,i_{|S_{k}|}$ enumerate the states in state space $S_{k}$ and $|S_{k}|$ denotes the number of states in $S_{k}$.
We define now the projection of $\hat{\pibd}^{(k)}(t)$ onto state space $S_{l}$ as follows
\begin{multline}
    \mathcal{P}_{k \to l} \hat{\pibd}^{(k)}(t) =\\ [\bar{P}^{(l)}(n_{j_1},o_{j_1},t),\ldots,\bar{P}^{(l)}(n_{j_{|S_{l}|}},o_{j_{|S_{l}|}},t)]^{\mathrm{T}},
\end{multline}
where the $j_1,\ldots,j_{|S_{l}|}$ enumerate the states in state space $S_{l}$, and
\begin{equation}
    \bar{P}^{(l)}(n_{j_m},o_{j_m},t) = \begin{cases}
        \hat{P}^{(k)}(n_{j_m},o_{j_m},t),&(n_{j_m},o_{j_m}) \in S_k,\\
        0,&\textrm{otherwise}.
    \end{cases}
\end{equation}

The proposed adaptive state reduction algorithm solves \eqref{eq:def:pibd_hat} and then maps the result to the reduced state space of the next interval in an iterative manner.
The complete algorithm is presented as Alg.~\ref{alg:cme_lumped_state_space} at the top of this page.

\begin{algorithm}[t]
    \caption{Iterative computation of $\pibd(t)$}
    \begin{algorithmic}[1]\label{alg:cme_lumped_state_space}
        \STATE \textbf{initialize:} $k=1$, $K=\left\lceil t/\Delta t \right\rceil$, $\hat{\pibd}^{(0)}(0) = \pibd_0$, $\Delta t$, $\epsilon$.
        \WHILE{$k \leq K$}
        \STATE Compute $S_k$ according to \labelcref{eq:def:Nmin,eq:def:Nmax,eq:def:Omin,eq:def:Omax,eq:def:S_k}.
        \STATE Set $\hat{\pibd}^{(k)}(t_{k}) = \mathcal{P}_{k-1 \to k} \hat{\pibd}^{(k-1)}(t_{k})$.
        \STATE Compute $\hat{\pibd}^{(k)}(t)$ for $t \in I_k$ by solving \eqref{eq:def:pibd_hat}.
        \STATE Set $k=k+1$.
        \ENDWHILE
        \STATE Return $\mathcal{P}_{K \to 0}\hat{\pibd}^{(K)}(t)$.
    \end{algorithmic}
\end{algorithm}

\section{Numerical Results}
\label{sec:numerical_results}
In this section, we present numerical results for the statistics of $N(t)$ and $O(t)$ obtained with Alg.~\ref{alg:cme_lumped_state_space} and compare these results with two reference models for $O(t)$, one reference model for $N(t)$, and stochastic \ac{PBS}.
The reference models for $O(t)$ are the statistical model based on the hypergeometric distribution proposed in \cite{lotter21}, denoted by $\mathcal{H}(o)$, and the binomial model obtained by assuming statistical independence of the receptors, $\mathcal{B}(o) = P_B(o;C,o(t)/C)$, where $P_B$ was defined in Section~\ref{sec:cme:state_reduction}.
In lack of any existing reference model for $N(t)$, we compare the predictions of our model for $P_N(n)$ with the binomial model obtained under the assumption that \acp{NT} are degraded independently of each other, i.e., $\mathcal{N}(n) = P_B(n;N_0,n(t)/N_0)$.
For the implementation details of the \ac{PBS}, we refer the reader to \cite{lotter20,lotter21}.
To compare the \ac{PBS} with the results obtained with Alg.~\ref{alg:cme_lumped_state_space}, we compute the empirical distribution of $N(t)$ and $O(t)$ at some time $t$ based on $6,000$ \ac{PBS} realizations.

For the numerical analysis, we consider three sets of parameter values, $\mathfrak{S}_0$, $\mathfrak{S}_1$, and $\mathfrak{S}_2$, listed in Table~\ref{tab:parameter_values}.
Further model parameters relevant for the \ac{PBS} and the state space model $\mathcal{S}$ but not for the \ac{CME} model considered in this paper are set according to \cite[Table~1]{lotter21}.

\begin{table}
    \vspace*{0.07in}
    \centering
    \caption{Parameter values for scenarios considered in Sec.~\ref{sec:numerical_results}.}
    \vspace*{-1mm}
    \footnotesize
    \begin{tabular}{| p{.18\linewidth} | r | r | r |}
        \hline & $\mathfrak{S}_0$\cite{lotter21} & $\mathfrak{S}_1$ & $\mathfrak{S}_2$\\ \hline
        $N_0$~$[-]$ & $\SI{1000}{}$ & $\SI{1000}{}$ & $\SI{250}{}$\\ \hline
        $C$~$[-]$ & $203$ & $600$ & $600$\\ \hline
        $\ka$~$[\si{\micro\meter\per\micro\second}]$ & $1.52 \times 10^{-5}$ & $4.48 \times 10^{-3}$ & $4.48 \times 10^{-4}$\\ \hline
        $\ke$~$[\si{\per\micro\second}]$ & $10^{-3}$ & $10^{-3}$ & $10^{-5}$ \\ \hline
        $\epsilon$~$[-]$ & $10^{-6}$ & $10^{-6}$ & $10^{-6}$ \\ \hline
        $\Delta t$~$[\si{\micro\second}]$ & $50$ & $50$ & $50$ \\ \hline
    \end{tabular}
    \label{tab:parameter_values}
\end{table}

$\mathfrak{S}_1$ is used to model synapses in which the competition of \acp{NT} for receptors is relatively small as compared to $\mathfrak{S}_0$ as it is the case in the neuromuscular junction where more receptors are present than in central synapses \cite{holmes95}.
In this case, the assumption underlying the model proposed in \cite{lotter21} is not fulfilled.
$\mathfrak{S}_2$ models a scenario in which many receptors compete for relatively few \acp{NT}.
Although \acp{NT} are usually more abundant than receptors, this situation may occur as a consequence of impaired vesicle loading \cite{pothos02}.

\subsection{Receptor Occupancy Statistics}

\begin{figure*}
    \centering
    \includegraphics[width=\linewidth]{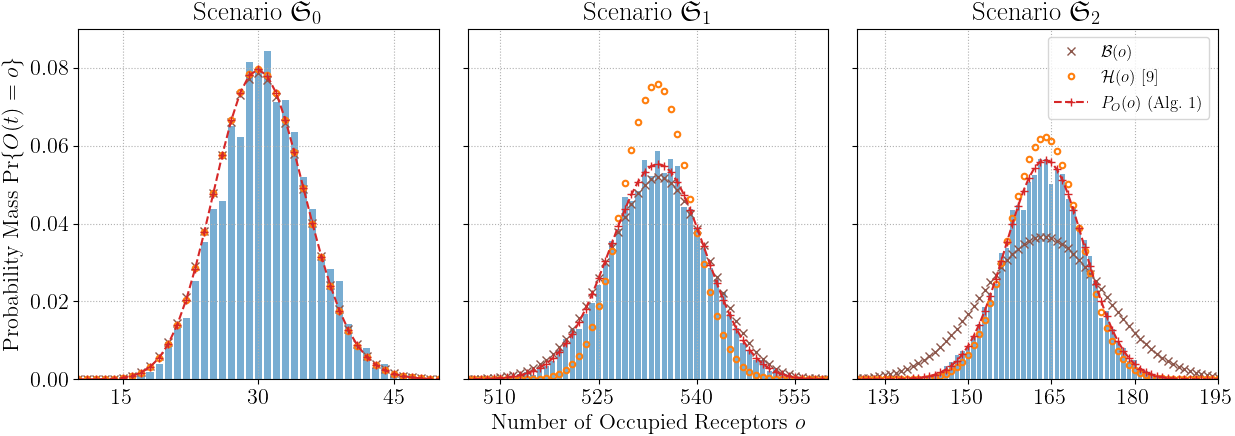}
    \caption{Probability mass function of $O(t)$ at $t=1$~$\si{\milli\second}$ as predicted by the model proposed in Sections~\ref{sec:system_model} and \ref{sec:cme} (red), the statistical model proposed in \cite{lotter21} (orange), and the binomial model (brown). Results from \acp{PBS} are shown in blue. The three subfigures correspond to the different scenarios $\mathfrak{S_0}$, $\mathfrak{S_1}$, and $\mathfrak{S_2}$, respectively, as defined in Table~\ref{tab:parameter_values}.}
    \label{fig:rec_occ_stats}
    \vspace*{-4.5mm}
\end{figure*}

Fig.~\ref{fig:rec_occ_stats} shows $\mathrm{P}_O(t)$ at $t=\SI{1}{\milli\second}$ as obtained by Alg.~\ref{alg:cme_lumped_state_space} and the reference models $\mathcal{H}$ and $\mathcal{B}$, as well as the results obtained by \ac{PBS}, for $\mathfrak{S}_0$, $\mathfrak{S}_1$, and $\mathfrak{S}_2$.
We observe from Fig.~\ref{fig:rec_occ_stats} that the model proposed in Sections~\ref{sec:system_model} and \ref{sec:cme} matches the empirical distribution obtained by \ac{PBS} accurately for all considered sets of parameters.
Also, both reference models $\mathcal{H}$ and $\mathcal{B}$ match the \ac{PBS} data for $\mathfrak{S}_0$.
However, $\mathcal{H}$ fails to reproduce $\mathrm{P}_O(t)$ for $\mathfrak{S}_1$.
The reason for this is that due to the large abundance of both \acp{NT} and receptors in $\mathfrak{S}_1$, there is neither competition of \acp{NT} for receptors nor competition of receptors for \acp{NT} and the main assumption for $\mathcal{H}$ is not fulfilled.
On the other hand, $\mathcal{B}$ fails to reproduce $\mathrm{P}_O(t)$ for $\mathfrak{S}_2$ the reason being that for $\mathfrak{S}_2$, the independence assumption underlying $\mathcal{B}$ is not fulfilled.
We conclude that the statistical model for $O(t)$ proposed in this paper is more robust towards parameter variations than previous models.

Finally, we observe from Fig.~\ref{fig:rec_occ_stats} that the variance of $O(t)$ and, consequently, the statistical dependence between the activation of different postsynaptic receptors depends largely on the choice of the synaptic parameters.
While correlation between receptors is rather strong in $\mathfrak{S}_2$, it is almost negligible in $\mathfrak{S}_0$ and $\mathfrak{S}_1$.
In $\mathfrak{S}_0$, on the other hand, competition of \acp{NT} for receptors is \mbox{stronger compared to $\mathfrak{S}_1$.}

\subsection{Neurotransmitter Degradation Statistics}

\begin{figure}
    \centering
    \includegraphics[width=.8\linewidth]{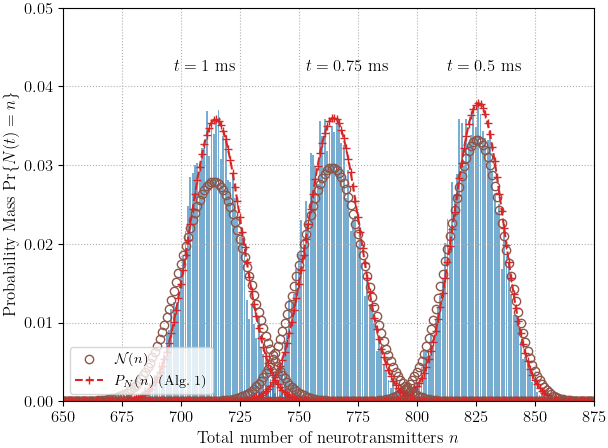}
    \caption{$P_N(t)$ at different time instants $t$ for $\mathfrak{S}_1$. The figure shows the \ac{CME} model proposed in this paper (red), the binomial model $\mathcal{N}$ (brown) and results from \acp{PBS} (blue).}
    \label{fig:inst_deg_stats}
\end{figure}

Now, we consider the statistics of $N(t)$.
Fig.~\ref{fig:inst_deg_stats} shows the statistics of $N(t)$ as obtained by Alg.~\ref{alg:cme_lumped_state_space}, reference model $\mathcal{N}$, and \ac{PBS} data at different time instants $t=\SI{0.5}{\milli\second},\SI{0.75}{\milli\second},\SI{1}{\milli\second}$ for parameter values $\mathfrak{S_1}$.
First, we observe from Fig.~\ref{fig:inst_deg_stats} that the results obtained by Alg.~\ref{alg:cme_lumped_state_space} match the empirical distribution of $N(t)$ very well for all considered time instants.
Furthermore, we observe from Fig.~\ref{fig:inst_deg_stats} that the degradation of single \acp{NT} is negatively correlated, since $P_N(t)$ as obtained by Alg.~\ref{alg:cme_lumped_state_space} is more concentrated compared to the binomial model $\mathcal{N}$.

From these results, we conclude that the proposed model can be used to gain novel insights into the impact of the various synaptic parameters on the statistics of synaptic signaling.

\vspace*{-3.3mm}
\section{Conclusion}
\label{sec:conclusion}
In this paper, we proposed a novel statistical model for the receptor occupancy and the neurotransmitter degradation in the synaptic \ac{DMC} system.
The proposed model is superior to existing models, because its applicability is not compromised by the simplifying assumptions underlying these models; neither does it require statistical independence of the receptors as the binomial model does, nor does it require competition of \acp{NT} for receptors or receptors for \acp{NT} as the model proposed in \cite{lotter21} does.
Furthermore, in contrast to existing models, the proposed model yields the {\em joint} distribution of the number of occupied receptors and the number of surviving \acp{NT} which fully characterizes the system state at each time instant.
The comparison with \ac{PBS} results underlines the high accuracy of the proposed model and the validity of the assumptions made to arrive at this model.
In summary, the proposed model provides a step forward towards a complete statistical characterization of the synaptic \ac{MC} system.

We conclude noting that the \ac{CME} model proposed in this paper can be used to study the depolarization of the postsynaptic membrane resulting from the stochastic activation of postsynaptic receptors \cite{rusakov20}.
However, due to space constraints, discussion and further investigation of this point is left for future work.

\appendix[Proof of Theorem \ref{thm:state_reduction}]
\label{sec:app:prf_thm_1}
From the structure of $\Abd(t)$, cf.~\eqref{eq:A:Q}, \eqref{eq:A:D}, we know that there is only probability flux from level $n+1$ to level $n$, not vice versa.
Hence, we conclude that
\begin{equation}
    \sum_{n=n_0}^{N} P_N(n,t+\Delta t) \leq \sum_{n=n_0}^{N} P_N(n,t),\label{eq:P_N:upper_tail}
\end{equation}
for any $n_0 \in \lbrace 0,\ldots,N \rbrace$, $\Delta t > 0$.
Eq.~\eqref{eq:P_N:upper_tail} provides an upper tail bound for $P_N(n,t+\Delta t)$ in terms of $P_N(n,t)$.
On the other hand, by the same argument
\begin{equation}
    \sum_{n=0}^{n_0} P_N(n,t) \leq \sum_{n=0}^{n_0} P_N(n,t+\Delta t).\label{eq:P_N:lower_tail}
\end{equation}
Let us consider the interval $I_k$.
By assumption, we know $P_N(n,t_k)$.
Then, with $N^{(k)}_{\mathrm{max}}$ as defined in \eqref{eq:def:Nmax}, we conclude from \eqref{eq:P_N:upper_tail} that $\sum_{n'=\Nmax}^{N} P_N(n',t) < \epsilon$ for any $t \in I_k$.
Now, let us consider the assumption that the \acp{NT} are degraded independently of each other.
Under this assumption, since all \acp{NT} are identical, $N(t)$ follows a binomial distribution with parameters $N_0$ and $\mathbb{E}[N(t)]/N_0 = n(t)/N_0$.
Indeed, this is a worst-case assumption with respect to the spread of $P_N(n,t)$, since in reality, the degradation of \acp{NT} is negatively correlated\footnote{To see the negative dependence of degradation events, consider one \ac{NT} $N_i$. The more \acp{NT} are degraded by time $t$, the more likely it is that $N_i$ binds to a free receptor and thus cannot be degraded by enzymes. On the other hand, the fewer \acp{NT} are degraded, the more \acp{NT} compete for receptors and it is less likely that $N_i$ finds a free receptor that prevents it from being degraded. This conclusion is also confirmed by the results presented in Fig.~\ref{fig:inst_deg_stats}.}, i.e.,
\begin{equation}
    \sum_{n=0}^{n_0} P_N(n,t) \leq \sum_{n=0}^{n_0} P_B(n;N_0,n(t)/N_0),\label{eq:P_N:lower_tail_binom}
\end{equation}
where $P_B(\cdot;n,p)$ as defined in Section~\ref{sec:cme:state_reduction} \cite{yu08}.
Now, with $\Nmin^{(k)}$ as defined in \eqref{eq:def:Nmin}, we conclude from \eqref{eq:P_N:lower_tail_binom} and \eqref{eq:P_N:lower_tail} that $\sum_{n'=0}^{\Nmin^{(k)}} P_N(n',t) < \epsilon$ for any $t \in I_k$.
Since the binding of \acp{NT} to receptors is also negatively correlated \cite{lotter21}, the upper and lower tail bounds for $O(t)$ follow from the same line of argumentation as \eqref{eq:P_N:lower_tail_binom}.
This concludes the proof.

\renewcommand{\baselinestretch}{0.91}
\bibliographystyle{IEEEtran}    
\bibliography{IEEEabrv,cme}
\end{document}